\documentclass[slac_one]{revtex4}
\usepackage{graphicx}
\usepackage{fancyhdr}

\usepackage{amsmath,amssymb,exscale}
\usepackage{psfrag,graphicx}
\usepackage[sub,ovp]{psfragx}
\usepackage{overpic}
\usepackage{epsfig}
\usepackage{color,amsmath,amscd,shadow,fancybox,axodraw}
\usepackage[latin1]{inputenc}

\newcommand{\Bt}{\tilde B}
\newcommand{\Wt}{\tilde W}

\newcommand{\Bslash}{\not \!\!B}
\newcommand{\Wslash}{\not \!\!W}
\newcommand{\Btslash}{\not \!\!\Bt}
\newcommand{\Wtslash}{\not \!\!\Wt}

\def\I{\rm 1\kern-.24em l}

\begin{document}

\title{Electroweak precision constraints on the \\ Lee-Wick Standard Model} %% Paper title goes here


\author{Ezequiel \'Alvarez}
\affiliation{CONICET and Departamento de F\'{\i}sica, FCEyN, Universidad de Buenos Aires,
Ciudad Universitaria, Pab.1, (1428) Buenos Aires, Argentina}
\author{Leandro Da Rold}
\affiliation{Instituto de F\'{\i}sica, Universidade de S\~ao Paulo,
Rua do Mat\~ao 187, S\~ao Paulo, SP 05508-900, Brazil}
\author{Carlos Schat}
\affiliation{CONICET and Departamento de F\'{\i}sica, FCEyN, Universidad de Buenos Aires,
Ciudad Universitaria, Pab.1, (1428) Buenos Aires, Argentina} 

\author{Alejandro Szynkman} 
\affiliation{Physique des Particules, Universit\'e
de Montr\'eal, C.P. 6128, succ. centre-ville, Montr\'eal, QC,
Canada H3C 3J7}

\begin{abstract}
We present the constraints on the parameter space of the Lee-Wick Standard
Model coming from electroweak precision observables. The model predicts a
large positive $S$ and a negative $T$. We show that it is possible to find some
regions in parameter space with a fermionic state as light as 2.4-3.5 TeV.
We also propose a simple extension of the model including a fourth
generation. In this case it is possible to pass the electroweak constraints
with Lee-Wick fermionic masses of order 0.4-1.5 TeV and Lee-Wick gauge
masses of order 3 TeV.
\end{abstract}

\maketitle

\thispagestyle{fancy}


\section{INTRODUCTION} % Section title should be in all capitals.

The Standard Model (SM) describes the electroweak (EW) interactions 
with an incredible precision. However, the instability of the Higgs
potential under radiative corrections signals our ignorance over the
real mechanism of electroweak symmetry breaking (EWSB) and has lead
to many extensions beyond the SM that attempt to solve the hierarchy problem. 
Recently, Grinstein, O'Connell and Wise proposed a new extension of
the SM~\cite{Grinstein:2007mp}, 
based  on the ideas of Lee and Wick \cite{Lee:1969fy,Lee:1970iw} for a
finite theory of quantum electrodynamics.
The building block of the Lee-Wick proposal is to consider that the
Pauli-Villars regulator describes a physical degree of freedom. In the
Lee-Wick Standard Model (LWSM), this idea is extended to all the SM
in such a way that the theory is free from quadratic divergences and 
the hierarchy problem is solved.

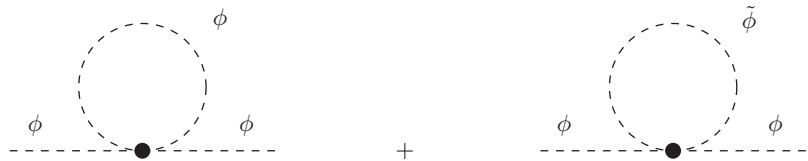
\begin{figure}
\begin{center}
\begin{picture} (500,80)(0,0)
	\Vertex(140,25){3}
	\Text(170,75)[c]{$\phi$}
	\Text(100,35)[c]{$\phi$}
	\Text(180,35)[c]{$\phi$}
	\DashLine(90,25)(140,25){3}
	\DashLine(140,25)(190,25){3}
	\DashCArc(140,49)(24,0,360){3}

	\Text(240,25)[c]{$+$}
	
	\Vertex(340,25){3}
	\Text(370,75)[c]{$\tilde \phi$}
	\Text(300,35)[c]{$\phi$}
	\Text(380,35)[c]{$\phi$}
	\DashLine(290,25)(340,25){3}
	\DashLine(340,25)(390,25){3}
	\DashCArc(340,49)(24,0,360){3}

\end{picture}
\caption{Cancellation of a quadratic divergence by a LW partner. }
\label{example1}
\end{center}
\end{figure}

The basic idea can be illustrated discussing the self energy of 
a scalar field, as shown in Fig.\ref{example1}. 
The propagators for the scalar field $\phi$ and its LW partner $\tilde \phi$
differ by an important overall sign
\begin{eqnarray}
D(p) = \frac{i}{p^2 - m^2} \qquad , \qquad 
{\tilde D}(p) = \frac{{-} i}{p^2 - M^2}  \ .
\end{eqnarray}
An explicit calculation shows that the quadratic divergence cancels and 
only a mild logarithmic divergence remains
\begin{eqnarray}
\Sigma(0) &=& i g \int \frac{d^4 p}{(2 \pi)^4} \frac{i}{p^2-m^2} {-} i g  \int \frac{d^4 p}{(2 \pi)^4} \frac{i}{p^2-M^2}
= i g \int \frac{d^4 p}{(2 \pi)^4} \frac{i (m^2 -M^2) }{(p^2-m^2)(p^2-M^2)} \ .
\end{eqnarray}
It is possible to understand the origin of the LW partner  $\tilde \phi$ 
starting from a higher derivative version of the scalar field lagrangian, 
as discussed in Ref.~\cite{Grinstein:2007mp}.

Similarly, in the LWSM every SM field has a LW-partner with
an associated LW-mass. These masses are the only new parameters in the
minimal LWSM. A potential problem in this model is that the LW-states
violate causality at the microscopic level due to the opposite sign of their
propagators. However, the authors of Ref.~\cite{Grinstein:2007mp} argued
that there is no causality violation on a macroscopic scale provided
that the LW-particles are heavy and decay to SM-particles.
The LWSM can also be thought as an effective theory coming from a higher derivative 
theory.

We refer the reader to
Refs.~\cite{Grinstein:2007mp,Alvarez:2008za} for more details and only quote here some
specific interaction terms that are useful to understand the
contributions to the EW observables ~\cite{Alvarez:2008za,Underwood:2008cr,Carone:2008bs}. The gauge-fermion interactions are:

\begin{eqnarray}\label{eqLint}
\mathcal{L}_{int}=&-&\sum_{\psi=q_L,u_R,d_R}\![g_1\bar\psi(\Bslash+\Btslash)\psi
+g_2\bar\psi(\Wslash+\Wtslash)\psi]
+\sum_{\psi=q,u,d}\left[g_1\bar{\tilde\psi}(\Bslash+\Btslash)\tilde\psi
+g_2\bar{\tilde\psi}(\Wslash+\Wtslash)\tilde\psi\right].
\end{eqnarray}
Note that the LW-fermions couple to the gauge fields with the opposite
sign compared with the SM-fermions.

Different to the SM chiral fermions, the LW-fermions combine into
Dirac spinors of masses $M_{q,u,d}$. 
We will assume that the LW-fermions transforming in the
same representation of the gauge symmetries have the same mass, which
is compatible with the minimal flavor violation principle. 

\section{TREE LEVEL CONTRIBUTIONS TO $S$ AND $T$}\label{tree}

In the LWSM the mixings between the gauge bosons and their
LW partners induce non-canonical couplings with
the SM fermions, as noticed in Ref.~\cite{Alvarez:2008za}.
A shift in the gauge fermion couplings can
be reabsorbed into the oblique parameters. Therefore, to correctly define
the oblique parameters $S,T$ it is necessary to properly normalize
the couplings between the fermions and the gauge bosons. This
observation was overlooked in Ref.~\cite{Grinstein:2007mp}. We find it
useful to work in the effective theory obtained after integrating out
the heavy LW fields at tree level. 
The canonical couplings are obtained  after a field redefinition: 
\begin{eqnarray}
\mathcal{L}_{eff}& \supset &- (g_2 + \delta g_2 ) 
W^{\mu a}J_\mu^a
- (g_1 + \delta g_1)  B^{\mu}J_\mu^Y  
=- g_2 {\bf W}_{\bf SM}^{\mu a}J_\mu^a-g_1
{\bf B}_{\bf SM}^{\mu}J_\mu^Y \; . 
\end{eqnarray}
where $J_\mu^a$ are the usual currents of SM fermions and $\bf W_{SM}$ and $\bf B_{SM}$ are identified as the SM fields. 

The effective Lagrangian written in terms of the SM fields 
\begin{eqnarray}
{\cal L}_{eff}  \supset -\frac12 {\Pi_{3B}'(0)} {\bf W}_{\bf SM}^{\mu\nu} {\bf B}^{\bf SM}_{\mu\nu} + \frac12 g_2^2 \Pi_{33}(0) ({\bf W}_{\bf SM}^3)^2 +  \frac12 g_2^2 \Pi_{11}(0) 
({\bf W}_{\bf SM}^1)^2  + \dots 
\end{eqnarray}
together with the usual definitions of the oblique parameters $S$ and $T$ 
\begin{equation}
S= \frac{16\pi}{g_1 g_2} \Pi_{3B}'(0) \; , \qquad
T=\frac{4\pi}{ s^2 c^2 m_Z^2}[\Pi_{11}(0)-\Pi_{33}(0)]\; , 
\end{equation}
lead to tree level contributions to
$S$ and $T$:
\begin{eqnarray}
S = 4 \pi v^2 \left( \frac{1}{M_1^2} + \frac{1}{M_2^2}
\right)+\mathcal{O}\left(\frac{v^4}{M_i^4}\right)\; , \qquad
T = \pi \frac{g_1^2 + g_2^2}{g_2^2} \frac{v^2}{M_1^2} \ , \label{Ttree}
\end{eqnarray}
where $v$ is the vacuum expectation value of the Higgs field and $M_{1,2}$ 
are the mass parameters of the LW fields $\tilde B, \tilde W$, respectively.
Notice that the sign difference between $T$ in 
Eq.~(\ref{Ttree}) and the result of Ref.~\cite{Grinstein:2007mp} is due to the additional contribution coming from the redefinition of the gauge fields mentioned above.
We can see that for $M_1\rightarrow \infty$ the tree level $T$ parameter
vanishes as expected, since in this limit we recover a custodial
symmetry in the LW-gauge sector.

\section{RESULTS INCLUDING ONE LOOP CONTRIBUTIONS TO $S$ AND $T$ }
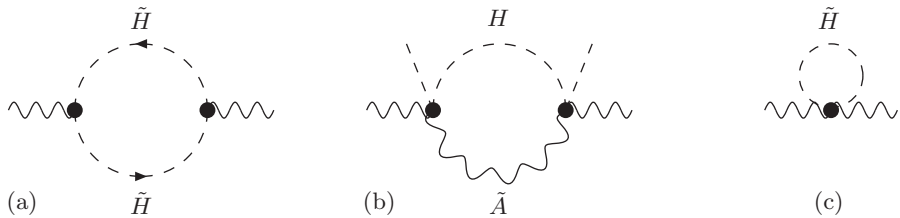
\begin{figure}[h]
\begin{center}
\begin{picture}(300,60)(0,0)
	\Vertex(-45,25){3}
	\Vertex(5,25){3}
	\Text(-20,60)[c]{$\tilde H$}
	\Text(-20,-10)[c]{$\tilde H$}
	\Photon(-70,25)(-45,25){3}{3}
	\Photon(5,25)(30,25){3}{3}
	\DashArrowArc(-20,25)(25,0,180){4}
	\DashArrowArc(-20,25)(25,180,360){4}
	\Text(-65,-10)[c]{(a)}
	\Vertex(90,25){3}
	\Vertex(140,25){3}
	\Text(115,60)[c]{$H$}
	\Text(115,-10)[c]{$\tilde A$}
	\Photon(65,25)(90,25){3}{3}
	\Photon(140,25)(165,25){3}{3}
	\DashCArc(115,25)(25,0,180){4}
	\PhotonArc(115,25)(25,180,360){3}{6}
	\DashLine(90,25)(80,50){4}
	\DashLine(140,25)(150,50){4}
	\Text(70,-10)[c]{(b)}
	\Vertex(240,25){3}
	\Text(240,60)[c]{$\tilde H$}
	\Photon(215,25)(240,25){3}{3}
	\Photon(240,25)(265,25){3}{3}
	\DashCArc(240,38)(12,0,360){4}
%	\DashCurve{(240,25)(230,40)(240,50)(250,40)(240,27)}{4}
	\Text(240,-10)[c]{(c)}
\end{picture}
\\[.5cm]
\caption{One-loop Feynman diagrams contributing to the oblique
parameters involving the Higgs sector.~~~~~~~~~~~~~~~~~~~~~~~~}
\label{figLWHiggs}
\end{center}
\end{figure}

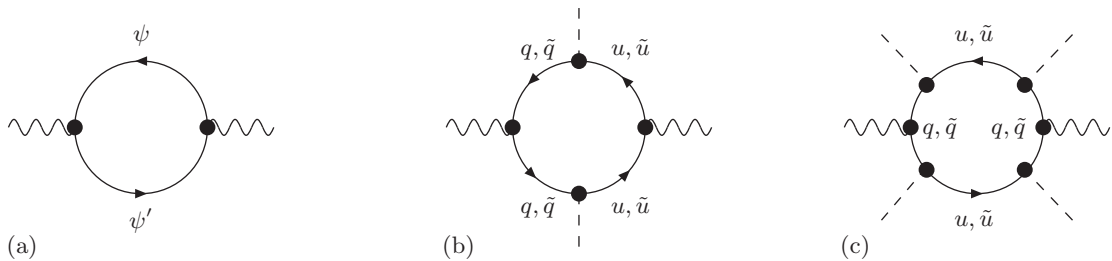
\begin{figure}[h]
\begin{center}
\begin{picture}(300,80)(0,0)

	\Vertex(-45,45){3}
	\Vertex(5,45){3}
	\Text(-20,80)[c]{$\psi$}
	\Text(-20,10)[c]{$\psi'$}
	\Photon(-70,45)(-45,45){3}{3}
	\Photon(5,45)(30,45){3}{3}
	\ArrowArc(-20,45)(25,0,180)
	\ArrowArc(-20,45)(25,180,360)
	\Text(-65,0)[c]{(a)}

	\Vertex(270,45){3}
	\Vertex(320,45){3}
	\Vertex(276,61){3}
	\Vertex(313,61){3}
	\Vertex(276,29){3}
	\Vertex(313,29){3}
	\Text(295,80)[c]{$u,\tilde u$}
	\Text(295,10)[c]{$u,\tilde u$}
	\Text(282,45)[c]{$q,\tilde q$}
	\Text(308,45)[c]{$q,\tilde q$}
	\Photon(245,45)(270,45){3}{3}
	\Photon(320,45)(345,45){3}{3}
	\ArrowArc(295,45)(25,0,180)
	\ArrowArc(295,45)(25,180,360)
	\DashLine(278,60)(259,80){4}
	\DashLine(312,60)(331,80){4}
	\DashLine(276,30)(259,10){4}
	\DashLine(312,30)(331,10){4}
	\Text(250,0)[c]{(c)}

	\Vertex(120,45){3}
	\Vertex(170,45){3}
	\Vertex(145,70){3}
	\Vertex(145,20){3}
	\Photon(95,45)(120,45){3}{3}
	\Photon(170,45)(195,45){3}{3}
	\ArrowArc(145,45)(25,0,90)
	\ArrowArc(145,45)(25,90,180)
	\ArrowArc(145,45)(25,180,270)
	\ArrowArc(145,45)(25,270,360)
	\DashLine(145,70)(145,90){4}
	\DashLine(145,20)(145,0){4}
	\Text(130,75)[c]{$q,\tilde q$}
	\Text(130,15)[c]{$q,\tilde q$}
	\Text(165,75)[c]{$u,\tilde u$}
	\Text(165,15)[c]{$u,\tilde u$}
	\Text(100,0)[c]{(b)}

\end{picture}
\caption{One-loop Feynman diagrams contributing to the oblique
parameters involving the fermionic sector. Diagram (a) is the
fermionic loop in the mass basis. Diagrams (b,c) are the first non-trivial
contributions from the up sector expanding in Yukawa mass insertions.}
\label{figLWfermions}
\end{center}
\end{figure}

The contributions from the Higgs sector come from the diagrams shown in Fig.\ref{figLWHiggs} and 
are found to be numerically small both for $S$ and $T$, when compared to 
their tree level values. 
 The fermionic contributions shown in Fig.\ref{figLWfermions}
are more important. The isospin splitting of the third generation gives a large negative 
contribution to the $T$ parameter that changes the sign of the tree level result. 
The resulting $T$ parameter is negative. On the other hand the fermionic contributions to $S$ are negative but small compared with the 
tree level value of $S$.

\begin{figure}[h]

$\begin{array}{ccc}
\begin{overpic}[width=7.5cm]{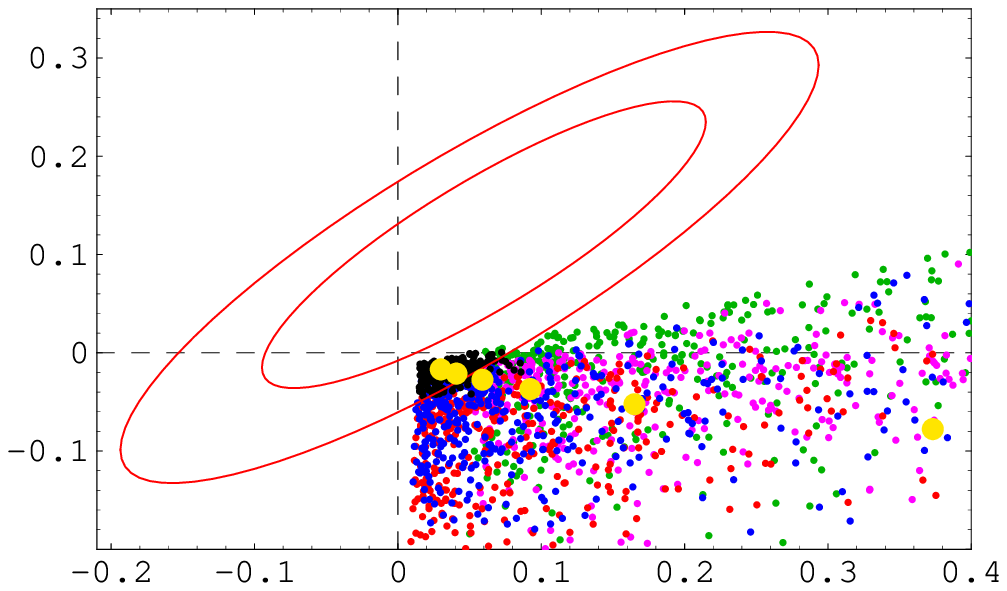}

\put(-7,30){\rotatebox{90}{$T$}}
\put(50,-5){$S$}

\end{overpic} & \ \ \ \ \ &
\begin{overpic}[width=7.5cm]{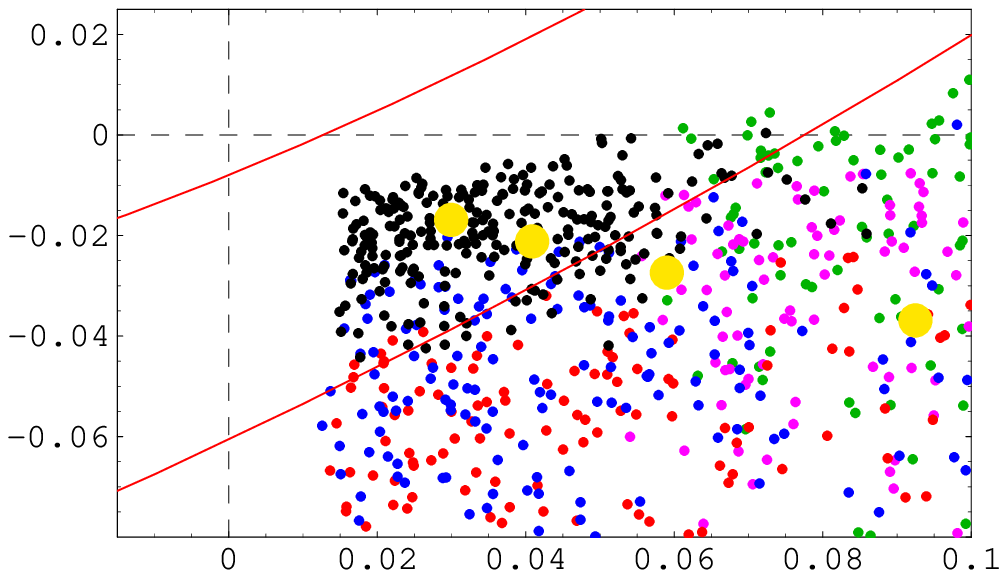}

\put(-7,30){\rotatebox{90}{$T$}}
\put(50,-5){$S$}

\end{overpic}
\end{array}$

\vspace{0.2cm}
\caption{\protect \small $68\%$ and $95\%$ Confidence Level contours
in the $(S,T)$ plane, and LWSM predictions. The black dots indicate
points where all four masses $M_1$ , $M_2$,  $M_q$  , $M_u$ are larger
than 4 TeV. Coloured points correspond to cases where at least one mass
is less than 4 TeV. The colour indicates which mass 
is below 4 TeV: green, magenta, red, blue dots correspond to 
$M_1$ , $M_2$,  $M_q$  , $M_u$ less than 4 TeV, respectively. 
The yellow dots correspond to taking all masses equal
and 7,6,5,4 ... TeV, from left to right. 
\label{planeST}}
\end{figure}

The result of scanning the parameter space is shown in Fig.\ref{planeST}.
A careful examination of the numerical values spanned by the dots that 
fall into the ellipse allows to find bounds on the LW masses, see Ref.~\cite{Alvarez:2008za} for details.\\[.5cm]

\begin{figure}[h]
$\begin{array}{ccc}
\begin{overpic}[width=5.5cm]{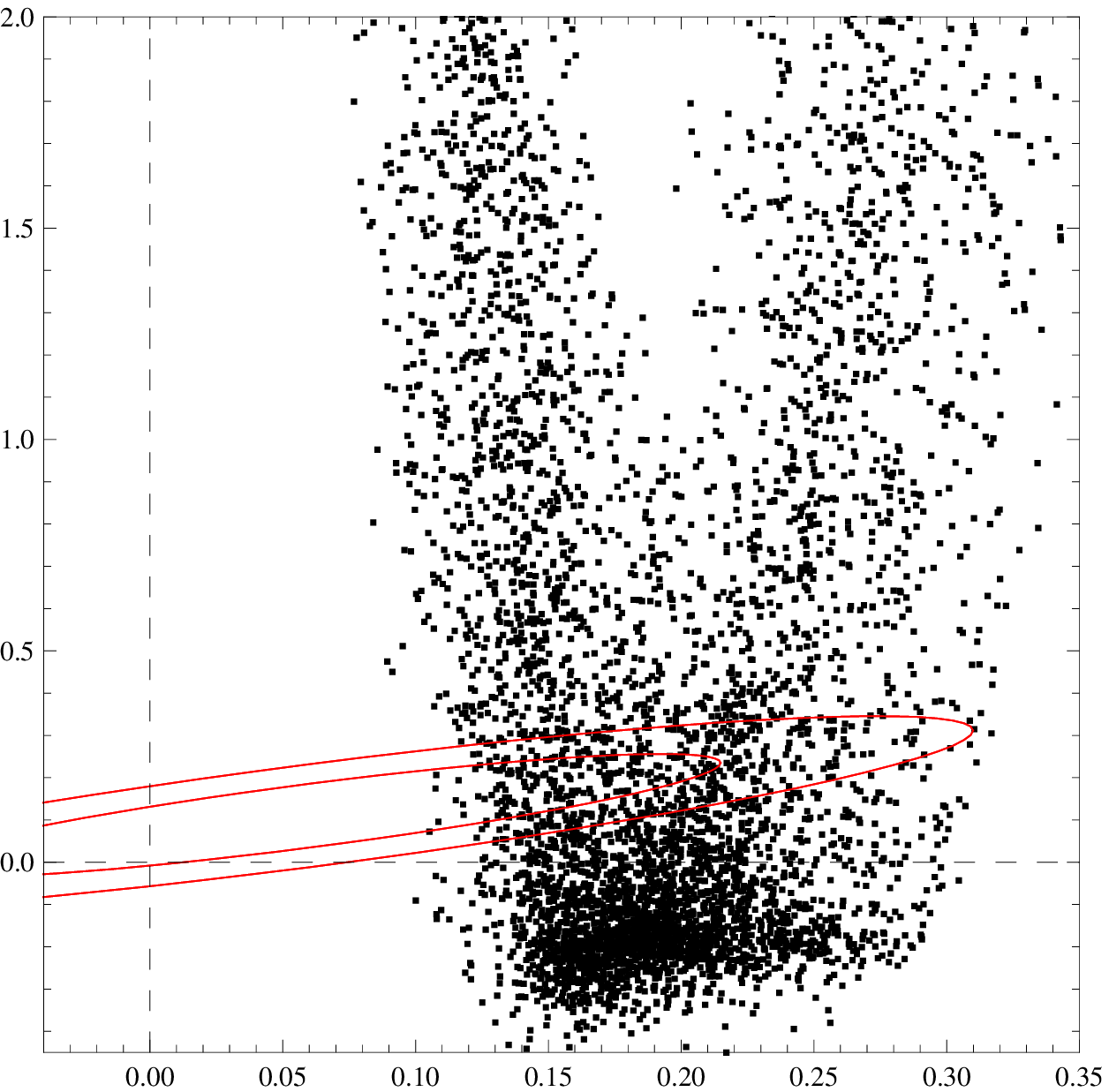}
\put(-7,50){\rotatebox{90}{$T$}}
\put(50,-5){$S$}
\end{overpic}& \ \ \ \ \ &
\begin{overpic}[width=5.5cm]{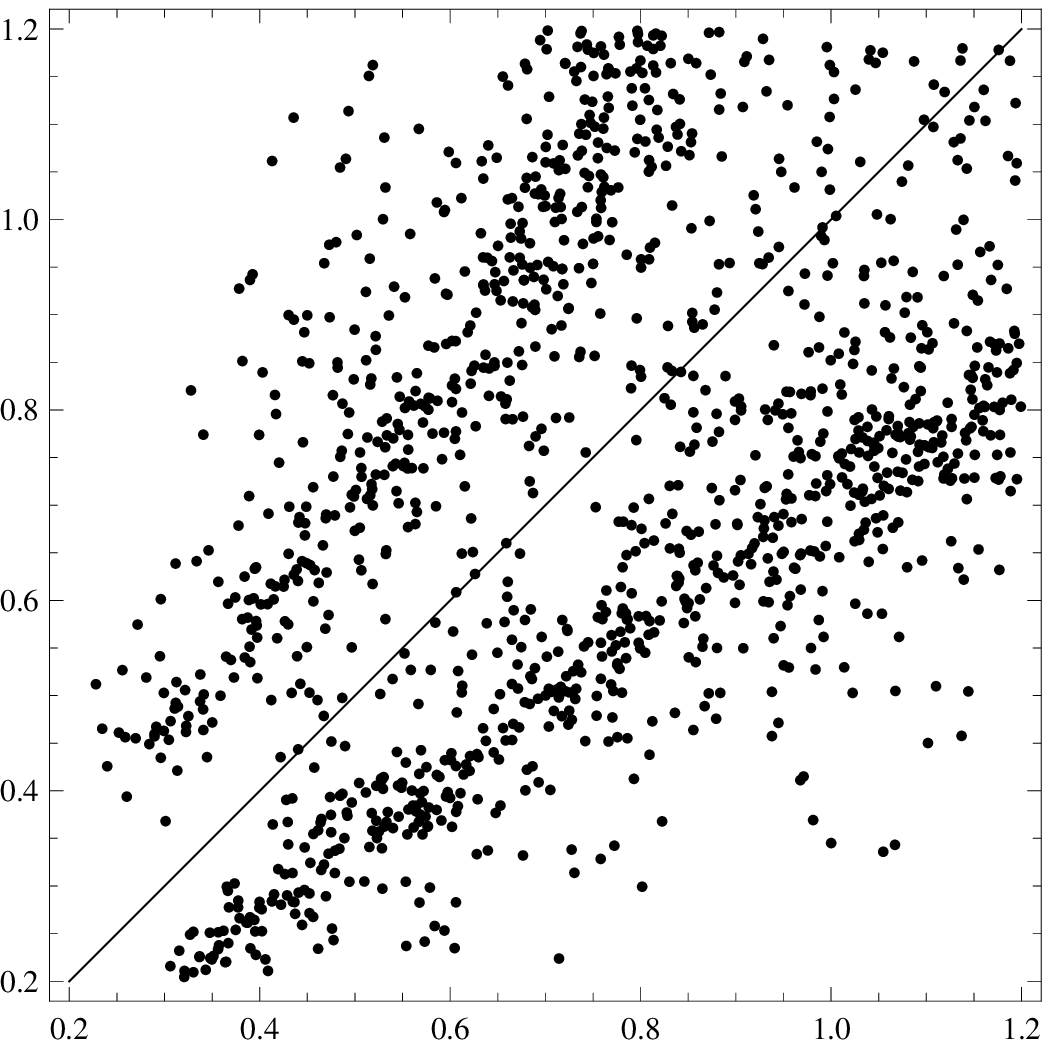}
\put(-10,40){\rotatebox{90}{$m_d^4$ [TeV]}}
\put(50,-7){$m_u^4$ [TeV]}
\end{overpic}
\end{array}$
\vspace{0.2cm}
\caption{$68\%$ and $95\%$ Confidence Level contours
in the $(S,T)$ plane, and predictions of the LWSM with a fourth
generation. The vector LW-masses are fixed to 3 TeV and $M_q\simeq
1.5$ TeV. The Yukawas and fermionic LW-masses take values in the range
$0.2-1.5$ TeV.}
\label{fig4ta}
\end{figure}

The figures show that a scan of the parameter 
space gives $S$ and $T$ values in the lower right quadrant, which is the 
less favorable one. This translates into higher bounds for the LW masses. 
A positive contribution to $T$ would bring the dots to the more 
favorable 
upper right quadrant and therefore lower the bounds on the LW masses.
This can be achieved by extending the LWSM adding a fourth generation. The contribution to $T$ is proportional to the 
isospin breaking of the new generation, which allows for more freedom, in 
contrast to the third generation, where the isospin breaking is fixed 
by the experimentally known masses. The new contribution to $S$ remains small
and the results are shown in Fig.\ref{fig4ta}. For a more thorough
discussion we refer the reader to Ref.~\cite{Alvarez:2008za}.

\section{CONCLUSIONS}
We discussed the electroweak precision constraints for the LWSM \cite{Alvarez:2008za}. We find sizeable tree level contributions for $S$ and $T$. After including the 
one loop contributions and scanning the parameter space we 
see that the LWSM is mostly located in the  $S>0 , T<0$ quadrant. A careful 
analysis of the ($S$,$T$) values compatible with experimental data leads to 
bounds that imply LW-fermion masses greater than 2.4 -3.5~TeV  
and LW-gauge bosons masses  greater than 5-8~TeV.   
Extending the LWSM adding a 4th-generation allows to relax these bounds. In 
this case we find that LW-fermion masses as small as 0.4 -1.5~TeV 
and LW-gauge   bosons masses as small as 3~TeV are still compatible 
with electroweak precision data.

\end{document}